\documentclass[aps,prl,twocolumn,groupedaddress]{revtex4}
\usepackage[dvips]{graphicx}
\usepackage{bm}

\begin{document}

\title{Quantum and classical localization in the lowest Landau level}
\author{Nancy Sandler}
\altaffiliation{Dept. of Physics and Astronomy, Ohio University, Ohio,
OH 45701}
\author{Hamid R. Maei}
\author{Jan\'{e} Kondev}
\affiliation{Department of Physics, Brandeis University ,Waltham, MA 02454}

\date{\today}

\begin{abstract}
Spatial correlations of occupation probabilities, if their decay is
not too fast, can change the critical exponents for classical percolation.
From numerical studies of electron dynamics in the lowest
Landau level (LLL) we demonstrate the quantum analog of this effect.
Similar to classical percolation, we find that the extended Harris
criterion applies to localization in the LLL. These results
suggest experiments that might probe new quantum critical points in
the integer quantum Hall setting.

\end{abstract}
\maketitle

Quantum or classical dynamics can lead to localization of a particle
moving in a random environment. The localized state in the quantum
case is described by a wave-function which spreads over a finite
distance, the localization length. In the classical case particle
trajectories are closed orbits limited in extent.  The mechanism by
which localization can occur in the two settings is very different as
quantum interference, which plays a crucial role in quantum
localization, has no classical counterpart. In this letter we report
on the intriguing relation between classical and quantum localization
in the setting of the integer quantum Hall (IQH) effect, when a
power-law correlated disorder potential is present.

The problem of localization in the IQH setting has been studied since
the effect was first discovered. Localization is believed to be
responsible for the plateaus in the Hall resistance which are
observed as a function of the applied magnetic field
\cite{ref:Prange}.  Transitions between adjacent plateaus have been
identified with a zero-temperature quantum critical point
\cite{ref:Levine} which is characterized by the divergence of the
localization length as the magnetic field is tuned to its
critical value.
Support for this picture is provided by observations of critical
scaling in experiments of Wei {\em et al.}\cite{ref:Tsui} and Koch
{\em et al.}\cite{ref:Koch}, which found a localization length
exponent $\nu_q \approx 2.3$.

Theoretical studies of localization in the IQH system usually take the
semiclassical picture \cite{ref:Trugman} as their starting point. Here
the electron's motion is split into the fast cyclotron rotation and
the slow ${\bf E}\times{\bf B}$ drift of the guiding center along
lines of constant electrostatic potential. The potential $V({\bf r})$,
caused by impurities, is assumed to be random. Quantum mechanics
enters at the saddle points of the potential where the electron can
tunnel from one equipotential orbit to another, while the
perpendicular magnetic field results in a random Aharonov-Bohm phase
between two tunnelling events.  A lattice model which describes the
network of saddle points was introduced by Chalker and Coddington
\cite{ref:Chalker}, and it leads to a value of the localization
exponent consistent with experiments.

The classical limit of the network model corresponds to percolation
\cite{ref:Lee}. The classical trajectories are equipotential lines of
$V({\bf r})$, which can be mapped to the hulls of percolation clusters
\cite{ref:Isichenko}.  However, little is known about quantum localization
when the classical limit is described by {\em correlated percolation}.
Namely, as shown by Weinrib \cite{ref:Weinrib2},
power-law correlations of the random potential can change the
percolation critical point leading to a different fractal geometry of
its equipotential
lines. This happens when the exponent $\alpha$, which characterizes the
spatial decay of the potential correlations (see Eq.~\ref{eq:potcorr}),
satisfies $\alpha < 2/\nu_c$; $\nu_c=4/3$ is the correlation length
exponent for percolation \cite{ref:Dennijs}. Here we
investigate the effect that changing the fractal geometry of classical
trajectories, has on quantum localization in the IQH system. We find new
localization critical points when $\alpha < 2/\nu_q$, i.e., the quantum
analog of correlated percolation. (This problem was previously studied by
Cain {\em et al.}~using real space renormalization group techniques, but no
definite proof of the effect was found \cite{ref:Cain}.)

\paragraph{Quantum dynamics in the lowest Landau level}

The Hamiltonian for two-dimensional spinless electrons confined to the
x-y plane, and under the combined effects of a magnetic field ${\bf B}
= B \hat{z}$ and a random potential $V({\bf r})$, is
$H = 1/2m ({\bf p}-e{\bf A})^2 + \sum_{{\bf k}}
V_{-{\bf k}}\, \rho_{{\bf k}}$. Here ${\bf A}$ is the vector
potential, $\rho_{{\bf k}} = e^{i{\bf k}\cdot{\bf r}}$ is the
one-particle density operator, and $V_{{\bf k}}$ is the Fourier
transform of the disorder potential.  At high magnetic fields (or low
temperatures), the quantum dynamics of the electron are governed by the
projection of the Hamiltonian onto the lowest Landau level (LLL),
\begin{equation}
\hat{H}\, =\, \sum_{{\bf k}}\, V_{-{\bf k}}\,\hat{\rho_{{\bf k}}} ,
\label{eq:hamilt}
\end{equation}
where $\hat{\rho}$ is the projected density operator. We focus
our attention on the localization properties of the eigen-functions of
$\hat{H}$. Previous numerical studies \cite{ref:Huckestein1} indicate
that the localization length increases towards the center of the LLL
band, with a localization length exponent $\nu_{q}\approx 7/3$.  In
these studies the random potential was assumed to be short range
correlated.  Here we take up the question of how power-law correlated
potentials might affect the value of $\nu_q$.

Recently, Sinova {\em et al.}~\cite{ref:Sinova} have shown that the
localization length exponent can be computed from the disorder
averaged density-density correlation function projected onto the
LLL. It was further shown by Gurarie and Zee \cite{ref:Gurarie} that
the classical limit of the time evolution of this correlation function
describes electron drift along the equipotentials of $V({\bf r})$, as
expected. These results lead to a simple scaling argument
for the subdiffusive spreading of wavepackets in the lowest Landau
level \cite{ref:Boldyrev}.  Namely, start by assuming that the disorder potential
completely breaks the degeneracy of the LLL, leaving only one extended
state at energy $E_c=0$.  Let us examine the time evolution of a
wavepacket $\left|\psi_{E}(t)\right>$ constructed from localized
eigenstates of Eq.~(\ref{eq:hamilt}) that are taken from an interval
of width $\Delta$ centered around $E$.  The dispersion of this
wavepacket, $<\Delta x^2(t)>_{E} \equiv\left<\psi_{E}(t)\left|
x^2\right|\psi_{E}(t)\right>$, is expected to be diffusive for short
times ($t \ll \xi^2(E)/D$, $D$= diffusion constant), attaining a
constant value set by the localization length $\xi(E)$ at long
times. This can be summarized by a scaling form:
\begin{equation}
\overline{<\Delta x^2(t)>_{E}}\,=\,D\,t\, f\!
\left(\frac{D\,t}{\xi^2(E)}\right)
\label{eq:defofg}
\end{equation}
where the bar denotes disorder averaging and
$f$ is a scaling function, with properties: $f(x)\to {\rm const}$
for $x\ll 1$, and $f(x)\to {\rm const}/x$ for $x\gg 1$. Now, if we
construct a wave-function in the LLL as a superposition of the wavepackets
$\left|\psi_{E}(t)\right>$ its dispersion can be written as
\begin{equation}
\overline{<\Delta x^2(t)>} \simeq \sum_{E\to -\infty}^{E=0}
\overline{<\Delta x^2(t)>_{E}} ,
\end{equation}
where the contributions of the non-diagonal terms can be neglected since
only the state at $E=0$ is extended.  Replacing $\overline{<\Delta
x^2(t)>_{E}}$ by Eq.~(\ref{eq:defofg}), making use of $\xi(E) \sim
E^{-\nu_q}$, and the non-singular nature of the density of states for
$E\to 0$, we obtain:
\begin{equation}
\label{eq:Xsq}
\overline{<\Delta x^2(t)>} \sim (D\;t)^{1-1/2\nu_q} \ .
\end{equation}
Therefore, the spread of the wavefunction at late times is
subdiffusive due to the presence of the extended eigenfunction at the
band center, and the anomalous diffusion exponent
$\theta=1-1/2\nu_q$. This forms the basis of the numerical method we
use below to compute $\nu_q$.

\paragraph{Power-law correlated disorder potential}

To study quantum dynamics and localization in the LLL in the presence of
power-law correlated disorder potentials, we make use of the Hamiltonian in
Eq.~(\ref{eq:hamilt}) with  $V({\bf k})$ a random Gaussian variable with
zero mean and variance given by $|{\bf k}|^{\alpha-2}$. After Fourier
transforming this leads to a potential which is power-law correlated in real
space:
\begin{equation}
\left<\,V({\bf r}) V({\bf 0})\,\right> \propto \frac{1}{|{\bf r}|^{\alpha}} \ .
\label{eq:potcorr}
\end{equation}
The projected Hamiltonian matrix is written in the basis of states
given by the Landau eigenstates on the torus, which in the
Landau gauge can be written as:
\begin{eqnarray}
\psi_{l}(x,y)&=&\sum_{m=-\infty}^{\infty}\exp\left(2\pi i
\frac{y}{L}(N\,m+l) \right) \nonumber \\
&&\times
\exp\left(-\pi\frac{N}{L^2}[x - \frac{L}{N}(l + N\,m)]^2\right) \ .
\label{eq:psialpha}
\end{eqnarray}
Here $l$ goes from $0$ to $N-1$, labelling the $N$ states on the
torus, $N$ is the number of flux quanta through the torus, and $L$ is
the system size.  These wavefunctions are defined on
$[0,L)\times[0,L)$, they are centered at $x = Ll/N$ and have
appreciable amplitude in a narrow strip of width given by the magnetic
length {$\ell_c = \sqrt{L^2/2\pi N}$}.
(Below we set $L=1$, rendering $(x,y)$ dimensionless.)

The electron density operator $\hat{\rho}(k_1,k_2) = \mbox{exp}(2\pi i
(k_1x+k_2y))$ projected onto the LLL is a matrix of the form:
\begin{equation}
\hat{\rho}(k_1,k_2)\,=\,\mbox{exp}
\left(-\frac{k_1^2\,+\,k_2^2}{2\,N}\pi\right) L(k_1,k_2) ,
\label{eq:rohat}
\end{equation}
where $k_1, k_2$ are integers in the units chosen, ($(k_1, k_2) \neq
(0,0)$), and the matrix elements of $L(k_1,k_2)$ are given by:
\begin{equation}
\left[L(k_1,k_2)\right]_{l_1, l_2} = e^{2\pi i[k_1\,k_2/2 + k_1(l_1-1)]/N}
 \,\delta_{l_1, l_2 -k_2}\,|_{mod N} \ .
\label{eq:Lk1k2}
\end{equation}
(The formalism used to project the density operator $\rho_{{\bf k}}$
onto the LLL was developed in Ref.~\onlinecite{ref:Girvin1}.)

\begin{figure*}[t]
\includegraphics[scale=1,width=6.9in,height=2.1in,keepaspectratio=false]{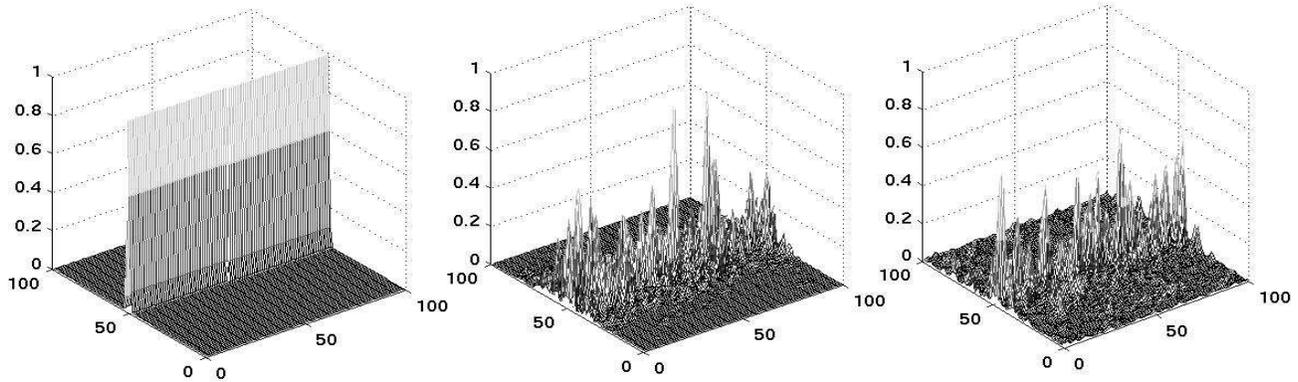}
\caption{Evolution of a wavepacket composed of
  all eigenstates and initially localized at $x=0.5$. The real
  function plotted is the square of the amplitude of the wavepacket.\vspace{0.4in} }
\label{fig1:wavepack}
\end{figure*}

For a given realization of $V({\bf r})$, the eigenvectors and
eigenvalues of $\hat{H}$ are obtained by exact diagonalization
\cite{ref:Numrec}. A wave-packet localized along $x$ is made using all
the eigenstates. As it evolves it spreads, and its width in the
x-direction is computed as a function of time. For one realization of
the short-range correlated random potential, Fig.\ref{fig1:wavepack}
shows a set of snapshots for the evolution of a wavepacket initially
localized at $x = 0.5$. To obtain $\overline{<\Delta x^2(t)>}$ the
width of the wavepacket is averaged over all initial positions and
over 1000 different disorder realizations; Eq.~\ref{eq:Xsq} is then
used for computing $\nu_q$.

\begin{figure}
\includegraphics[scale=1,draft=false,width=5.1in,
height=2.3in,keepaspectratio=true]{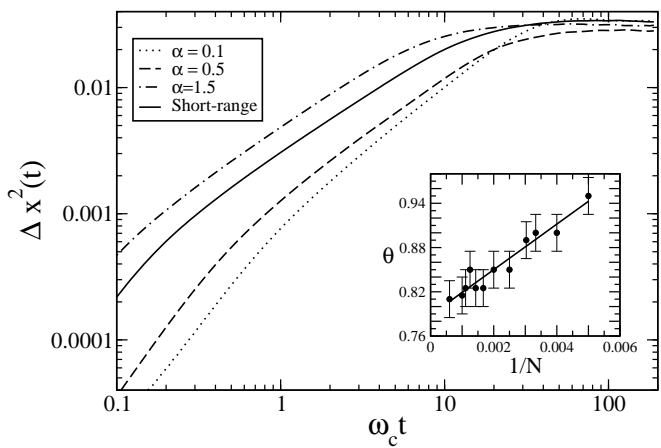}
\caption{\label{fig2:quanlr} Disorder averaged spread of the
  wave-packet, for $N=1000$ basis states in the LLL, as a function of
  time, and for different disorder potentials.  Anomalous diffusion is
  evident at intermediate times.  Inset: The exponent $\theta$ for
  different values of $N$ in the case of a short-range correlated
  disorder potential.}
\end{figure}

First, we checked this numerical procedure for a short-range
correlated disorder potential, where the Fourier components of $V({\bf
r})$ (in units of $\hbar \omega_c$) are independent and uniformly
distributed on the interval $[-0.5, 0.5]$.  The value of
$\theta=1-1/2\nu_q$ is given by the slope of $\overline{<\Delta
x^2(t)>}$ when plotted on a log-log graph.  To take into account finite size
effects we repeat the computation of $\theta$ for $200\le
N\le 1500$ and plot it as a function of $1/N$; see inset of
Fig.~\ref{fig2:quanlr}. By linear extrapolation we find $\nu_{q} =
2.33(9)$ in the $N\to\infty$ limit, consistent with previous
results.  We then repeat the computation for long-range correlated
disorder potentials for different values of the parameter $\alpha$.
Results are shown in Fig.~\ref{fig2:quanlr}.

For each $\overline{<\Delta x^2(t)>}$ curve in Fig.~\ref{fig2:quanlr},
three regions can be identified.  Due to the finite system size, at
long enough times the spread of the wave-packet reaches a constant
value which corresponds roughly to $70\%$ of the system size. The
critical region corresponds to intermediate times, while the short
time behavior is described by the slope $\theta \approx 2$. Ballistic
motion ($\theta=2$) follows from a perturbative calculation of
$\overline{<\Delta x^2(t)>}$ at short times \cite{ref:SMKlong}.
Comparison among different $\overline{<\Delta x^2(t)>}$ curves shows
that the value of the slope in the critical region begins to increase
for $\alpha < \alpha^* \approx 0.75-0.80$.

\paragraph{Extended Harris criterion}

The effect of short-range correlated disorder on a critical point is
summarized by the Harris criterion \cite{ref:Harris}. The criterion
states that critical exponents for the disordered and the clean system
remain equal as long as the value of the correlation length exponent
$\nu$ satisfies: $d \nu - 2 \geq 0$, where $d$ is the dimensionality
of the system under consideration. It is derived by demanding that the
fluctuations of the random potential within a volume set by the
correlation length do not grow faster than its average value, as the
transition is approached.  An extension of the criterion was proposed
by Weinrib and Halperin \cite{ref:Weinrib1} to include power-law
correlated disorder potentials, like that in
Eq.~(\ref{eq:potcorr}). When applied to two-dimensional percolation
this "extended Harris criterion'' reads \cite{ref:Weinrib2}:

\begin{equation}
\alpha > {\alpha^{*}} \  \Rightarrow  \  \nu_{c}= \frac{2}{\alpha^*};
\ \ \ \
\alpha < {\alpha^{*}} \  \Rightarrow  \  \nu_{c}= \frac{2}{\alpha} \ .
\label{eq:clas_harris}
\end{equation}
where $\nu_c=4/3$ and $\alpha^* = 3/2$.
However, there is to our knowledge no
evidence of the validity of the extended Harris criterion for quantum
critical points. Since the localization transition in the LLL is
believed to be the quantum counterpart of the classical percolation
transition, it provides an ideal testing ground for the
applicability of the criterion in the quantum realm.

\begin{figure}
\includegraphics[scale=1.2,draft=false,width=6.5in,
height=2.2in,keepaspectratio=true]{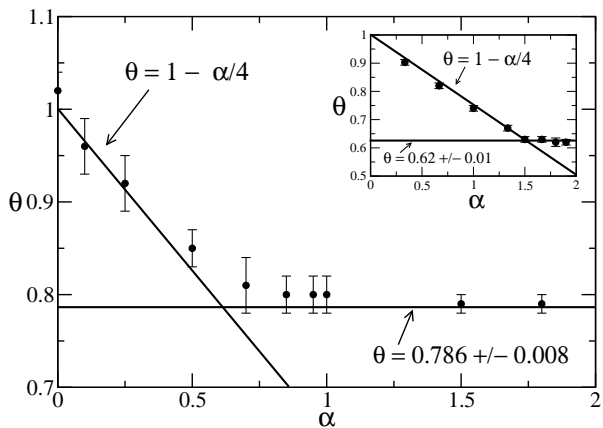}
\caption{\label{fig3:quant_nu_vsal} The anomalous diffusion exponent
$\theta=1-1/2\nu_q$ characterizes the spread of a wave-packet in the
LLL. The power $\alpha$ describes the long-range
nature of the correlations of the random potential. Full lines are
theoretical curves obtained from the extended Harris criterion. The
inset shows analogous results for the case of classical electron
motion \cite{ref:SMKlong}. }
\end{figure}

With this in mind we calculate the value of the anomalous diffusion
exponent $\theta=1-1/2\nu_q$  by computing
the value of $\theta$  in the critical region for different values for the
degeneracy of the LLL, ranging from $N=330$ to $N=1000$,
and then extrapolating to $N\to\infty$.  Results from this
computation, as well as
the theoretical prediction based on the extended Harris criterion are
shown in Fig.~\ref{fig3:quant_nu_vsal}. In the quantum case the
criterion reads as written in Eq.~(\ref{eq:clas_harris}) except now
$\alpha^*=2/\nu_q = 0.786(8)$.  In the inset of
Fig.~\ref{fig3:quant_nu_vsal} we also show results of numerical
analysis of classical electron motion in the same long-range
correlated random potential, and compare to the prediction from the
extended Harris criterion \cite{ref:SMKlong}.  The classical results
were reported previously in Ref.~\cite{ref:Prakash}.  Our data confirm
the validity of the extended Harris criterion for both quantum and
classical electron dynamics in the LLL.

The main result of this paper is contained in
Fig.~\ref{fig3:quant_nu_vsal}. From the striking resemblance between
classical and quantum behavior, we conclude that disorder correlations
affect in a qualitatively similar way quantum and classical percolation.
Indeed, the quantum version of the extended Harris criterion can be argued in
close analogy with its classical counterpart. In this case we should
consider the fluctuations of the random potential in a volume set by
the localization length $\xi(E)\sim E^{-\nu_q}$, and compare them to
$E$. The calculation that follows \cite{ref:SMKlong} exactly parallels that
done by
Weinrib and Halperin \cite{ref:Weinrib1}, and Eq.~(\ref{eq:clas_harris})
follows with $\nu_q$ replacing $\nu_c$.

An interesting consequence of the applicability of the extended Harris
criterion to the IQHT, is the conclusion that the simple relation
between the classical and quantum localization length exponent,
$\nu_q=\nu_c+1$ \cite{ref:Milnikov}, can not be valid in general.
According to the quantum version of Eq.~(\ref{eq:clas_harris}), for a
long-range correlated disorder with $0.8\lesssim \alpha \leq 1.5$ the
quantum localization exponent will be the same as in the case of
short-range correlated disorder while the classical one will vary
continuously with $\alpha$.

We conclude by speculating about a possible experimental test of
our findings. The random potential present in samples that exhibit the
IQH effect is thought to be short-range correlated, due to the nature
of the spatial distribution of impurities within the heterojunction
\cite{ref:Prange}. However, the effect of a random magnetic field in
the presence of a much stronger constant magnetic field is equivalent
to that of a random potential \cite{ref:Huckestein2}. This opens up
the possibility of engineering the properties of the disordered
environment seen by the electrons, by applying a random magnetic
field. This was, for example, demonstrated in
Ref.~\onlinecite{ref:Zielinski} using a magnetic material with a rough
contact surface placed in close proximity to the electron
layer. The height fluctuations of the material surface, and hence the
correlations in the random field, can in principle be controlled by
adjusting the growth conditions under which the material is
made. Magnetic decoration techniques or magnetic force microscopy
can be used to determine spatial correlations of the field and thus
extract the value of $\alpha$. At $T \simeq 25 mK$, the saturated
width of the Hall conductance step is $\Delta B \simeq 0.5 T$ for
$AlGaAs/GaAs$ structures. The temperature and $\Delta B$ place limits
on the size of the random field fluctuations, which should roughly be
in the interval $10^{-3}$T $\leq \delta B \leq 10^{-1}$T . Whether or
not magnetic materials with these properties can be prepared, and
placed in close proximity to the two-dimensional electron gas so as to effect
its dynamics, remains to be seen.

It is a pleasure to acknowledge useful conversations with B. Halperin,
J. Sinova, V. Gurarie, S. Boldyrev, J. Moore,  and S. Simon.  JK is
supported by the NSF under grant number DMR-9984471, and is a
Cottrell Scholar of Research Corporation.

\bibliography{RefQHT_jk}

\end{document}